\pdfoutput=1

\documentclass[prl,reprint,amssymb]{revtex4-2}

\usepackage{xcolor}

\usepackage{bm}
\usepackage{braket}
\usepackage{comment}
\usepackage{amsmath}  
\usepackage{amsfonts}
\usepackage{graphicx}
\usepackage{hyperref}
\hypersetup{
    colorlinks=true,
	citecolor=blue,
    linkcolor=red,
	urlcolor = blue
}

\begin{document}

\title{Magnon spectrum of altermagnets beyond linear spin wave theory: Magnon-magnon interactions via time-dependent matrix product states vs. atomistic spin dynamics}

\author{Federico Garcia-Gaitan}
\author{Ali Kefayati}
\author{John Q. Xiao}
\author{Branislav K. Nikoli\'c}
\email{bnikolic@udel.edu}
\affiliation{Department of Physics and Astronomy, University of Delaware, Newark, DE 19716, USA}


\begin{abstract}
The energy-momentum dispersion of magnons, as collective low-energy excitations of magnetic material, is standardly  computed from an effective quantum spin Hamiltonian but simplified via linearized Holstein-Primakoff transformations to describe noninteracting magnons. The dispersion produced by such linear spin wave theory (LSWT) is then plotted as ``sharp bands'' of infinitely long-lived quasiparticles. However, magnons are prone to many-body interactions with other quasiparticles---such as electrons, phonons or other magnons---which can lead to shifting (i.e., band renormalization) and broadening of ``sharp bands'' as the signature of finite quasiparticle lifetime. The magnon-magnon interactions can be particularly important in antiferromagnets (AFs), and, therefore, possibly in newly classified altermagnets sharing many features of  collinear AFs. Here, we employ nonperturbative  quantum many-body calculations, via time-dependent matrix product states (TDMPS), to obtain magnon spectral function for RuO$_2$ altermagnet whose effective quantum spin Hamiltonian is put onto 4-leg cylinder. Its upper band is shifted away from upper ``sharp band'' of LSWT, as well as broadened, which is explained as the consequence of {\em hybridization} of the latter with three-magnon continuum. This implies that two-magnon Raman scattering spectra {\em cannot} be computed from LSWT bands, which offers a litmus test for the relevance of magnon-magnon interactions.   Finally, we employ atomistic spin dynamics (ASD) simulations, based on classical Landau-Lifshitz-Gilbert (LLG) equation, to obtain magnon spectrum at finite temperature and/or at a fraction of the cost of TDMPS calculations. Despite including magnon-magnon interactions via nonlinearity of LLG equation, ASD simulations {\em cannot} match the TDMPS-computed magnon spectrum, thereby signaling {\em nonclassical} effects harbored by AFs and altermagnets. 
\end{abstract}

\maketitle

\textit{Introduction.}---The energy-momentum dispersion, standardly displayed~\cite{Kaxiras2019} as ``sharp bands''  along high-symmetry paths in the Brillouin zone, is a fundamental property of quasiparticle collective excitations in solids, such as electrons (in the sense of Landau quasielectrons), phonons and magnons~\cite{Kaxiras2019}. For example,  diagonalization of single-particle electron Hamiltonian---such as tight-binding~\cite{Konschuh2010}, or first-principles ones~\cite{Kogan2014} obtained from density functional theory (DFT)---yields bands whose sharpness signifies infinitely long-lived quasielectrons~\cite{Kaxiras2019}. However, in the presence of  Coulomb interaction effects that are beyond mean-field single particle approaches~\cite{Held2007}, one has to resort to many-body calculations  where quasielectrons  
decay and their bands broaden to quantify their inverse lifetime~\cite{Held2007, Watson2017}. Furthermore, when quasielectrons lose their identity completely, one obtains a  continuum of energies of the so-called Hubbard bands~\cite{Held2007, Watson2017}.

An analogous situation exists in the case of magnons--- quasiparticles introduced by Bloch~\cite{Bloch1930} as a wave-like disturbance in the local magnetic ordering of a magnetic material. Their  quanta~\cite{Bajpai2021} of frequency $\omega$ behave as bosonic quasiparticles carrying  energy $\hbar \omega$, spin $\hbar$ and quasimomentum $\hbar \mathbf{q}$. The ``sharp bands'', $\hbar \omega$ vs. $\mathbf{q}$ (as exemplified by red and orange lines in Fig.~\ref{fig:fig1}),  are standardly computed by extracting parameters of effective spin Hamiltonians~\cite{Szilva2023} from DFT~\cite{Smejkal2023}, which is then fed into Holstein-Primakoff (HP)~\cite{Primakoff1940} transformations that map original spin operators to bosonic ones. By linearizing HP transformations~\cite{Bajpai2021,Chudnovsky2006}, one obtains Hamiltonian quadratic in bosonic operators which is, therefore, exactly diagonalizable and provides the foundation of the so-called linear spin wave theory (LSWT)~\cite{Zhitomirsky2013,Gohlke2023,Habel2024}. However, experiments often measure~\cite{Li2016,Chen2022,Dai2000,Bayrakci2013} magnon damping, which is a problem of great interest for both basic research~\cite{Zhitomirsky2013} and applications in magnonics~\cite{Chumak2015,Chumak2022}. The origin of magnon damping can be traced to: magnon-magnon interactions, as described by second-quantized Hamiltonians containing products of three or more bosonic operators~\cite{Zhitomirsky2013,Bajpai2021}; magnon-phonon interactions~\cite{Dai2000}, especially relevant for recently discovered two-dimensional magnetic materials~\cite{Chen2022}; and magnon-electron interactions in magnetic metals~\cite{Buczek2009, TancogneDejean2020,  Hankiewicz2008, Tserkovnyak2009,Gallegos2024} or in  heterostructures involving metallic layers~\cite{Bertelli2021,ReyesOsorio2024c}. For example, the so-called Landau damping of magnons due to hybridization with the continuum of electronic Stoner excitations, where single-electron spin flips while transitioning from an occupied state with a given
spin projection to an empty state with an opposite spin projection, has been quantified using either random phase approximation applied to model Hamiltonians~\cite{Bonetti2022} or via first-principles methods like time-dependent DFT~\cite{Buczek2009,TancogneDejean2020}. Furthermore, magnon-magnon interactions can cause~\cite{Chernyshev2009,Zhitomirsky2013,Habel2024} magnon spontaneous decay, which broadens ``sharp bands'' of LSWT; magnon band renormalization shifting them; and a renormalization of one-magnon wavefunctions as hybrids of different LSWT band eigenstates. Such effects are often encountered in insulating antiferromagnets (AFs)~\cite{Bayrakci2013, Zhitomirsky2013,Harris1971}, particularly noncollinear ones~\cite{Chernyshev2009} or those hosting spin-orbit (SO)  coupling effects~\cite{Sourounis2024} and even down to zero temperature~\cite{Sourounis2024}; magnon topological insulators~\cite{Habel2024}; quantum paramagnets~\cite{Gallegos2024}; and quantum spin liquids~\cite{Smit2020, Winter2017}.

\begin{figure}
    \centering
    \includegraphics[width=\linewidth]{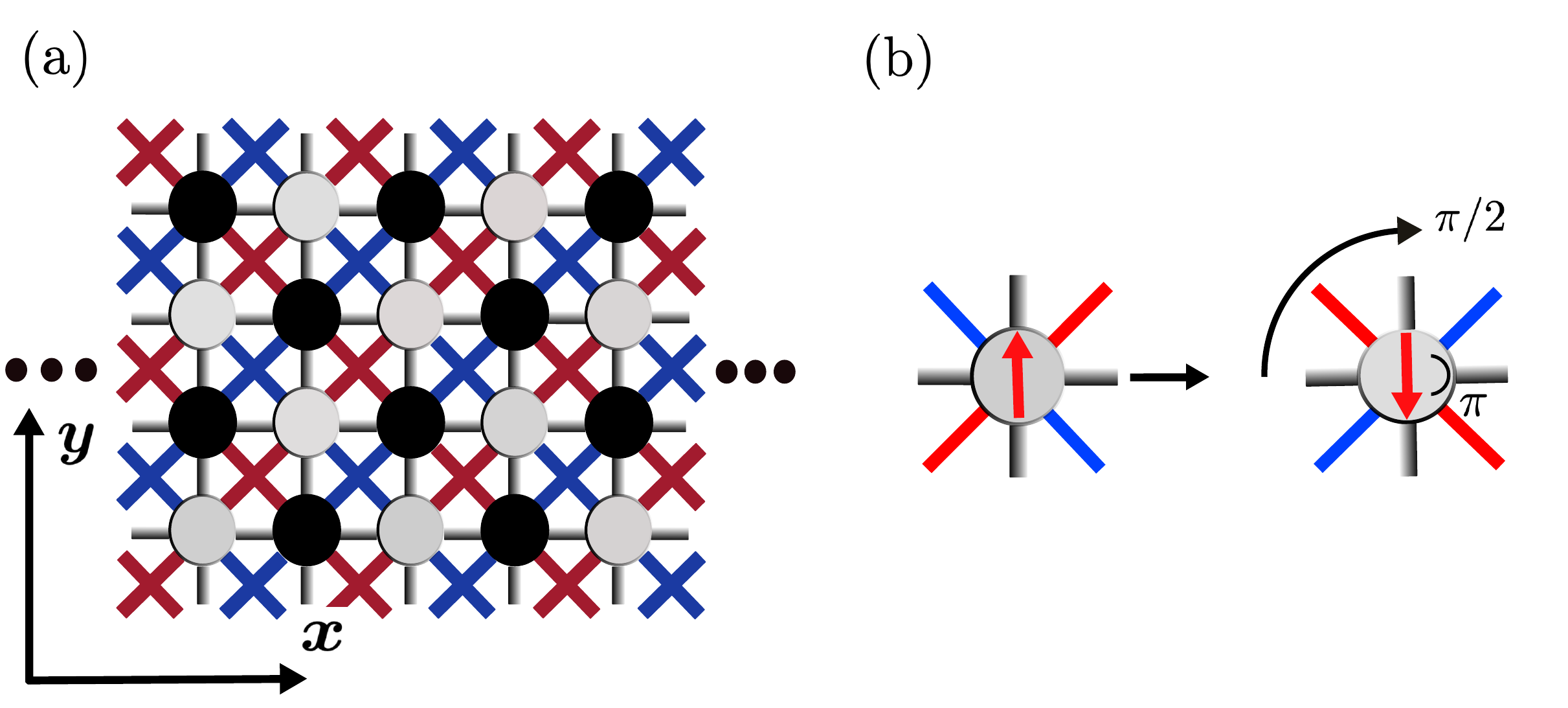}
     \caption{(a) Quasi-one-dimensional lattice, as a 4-leg cylinder because of periodic boundary conditions imposed along the transverse $y$-direction, onto which the effective spin Hamiltonian [Eq.~\eqref{eq:hamiltonian}] of  RuO$_2$~\cite{Smejkal2023} is placed for TDMPS calculations. Its three different Heisenberg exchange interactions are depicted using three different colors---red and blue for intrasublattice and gray for intersublattice. (b) Illustration of the symmetry group  connecting two sites (by $\pi/2$ rotation) or two spins (by $\pi$ rotation)  within two sublattices in (a).}
    \label{fig:fig0}
\end{figure}

\begin{figure*}[t!]
    \includegraphics[width=\linewidth]{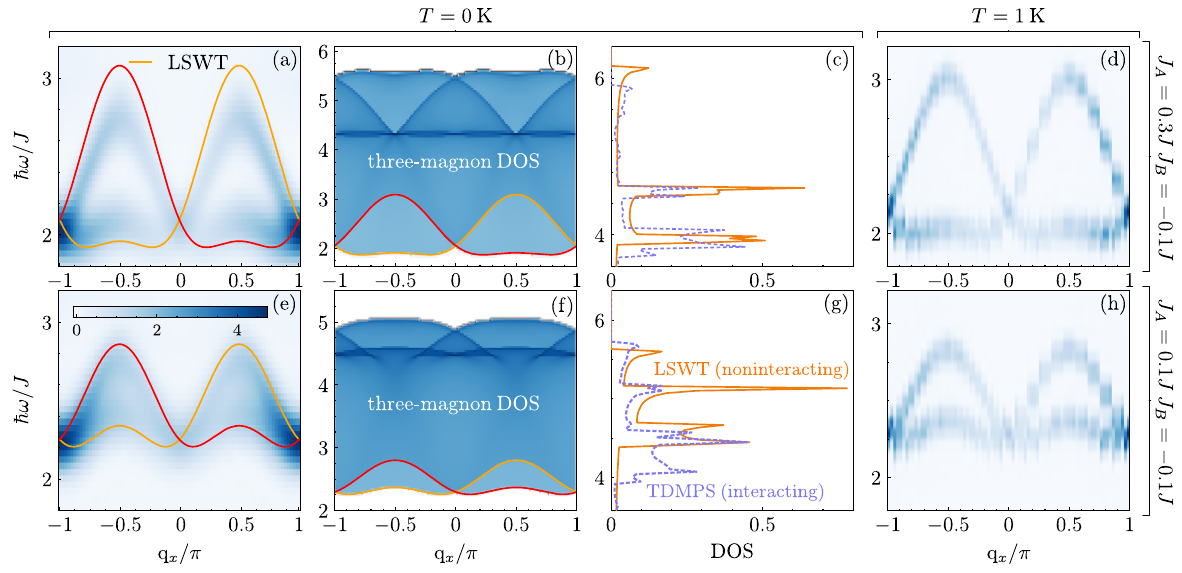}
    \caption{(a),(e) Magnon $\hbar \omega$ vs. $q_x$ spectrum, for $q_y=\pi/2$, of RuO$_2$ altermagnetic spin Hamiltonian [Eq.~\eqref{eq:hamiltonian}] put onto 4-leg cylinder in Fig.~\ref{fig:fig0}. The dispersion is obtained either from spectral function (blue trace) $A(q_x,\omega)$ [Eq.~\eqref{eq:spectralf}], as computed from quantum many-body calculations via TDMPS; or as ``sharp bands'' [red and orange lines plotting Eq.~\eqref{eq:HP}] computed from one-body calculations via LSWT. (b),(f) DOS of three-magnon continuum based~\cite{Winter2017} on one-magnon 
    ``sharp bands'' in (a),(e), respectively, is plotted together with those ``sharp bands'' in order to examine if one-magnon modes are energetically
     degenerate with the multi-magnon  continuum~\cite{Winter2017}. (c),(g) One-magnon DOS for LSWT vs. TDMPS bands, using doubled energy scales to allow for comparison~\cite{Prosnikov2021} with experimental two-magnon RS  spectra. (d),(h) The same information as in panels (a),(e) but obtained from classical ASD~\cite{Etz2015,Skubic2008} simulations of DSSF $\mathcal{S}^{zz}(q_x,\omega)$ [Eq.~\eqref{eq:dsf}] at \mbox{$T= 1$ K}. The difference between top and bottom row of  panels is that in the top row we use exchange interactions $J_A\neq J_B$ between quantum spins localized at NN lattice sites obtained from DFT calculations in Ref.~\cite{Smejkal2023}, while in the bottom row we artificially make them identical in absolute value. The color bar in panel (e) specifies values along blue traces in panels (a),(b),(d),(e),(f) and (h).}
    \label{fig:fig1}
\end{figure*}


Within quantum many-body theory, interaction induced shifting/renormalization of magnon energies $\hbar \omega(\mathbf{q})$ and decay-induced lifetime (i.e., broadening  of ``sharp bands'') can be rigorously defined using one-particle retarded Green's function (GF)~\cite{Chernyshev2009,Zhitomirsky2013,Habel2024,Bajpai2021}%
\begin{equation}\label{eq:gf}
G^{-1}(\mathbf{q},\omega) = \hbar\omega - \hbar\omega(\mathbf{q}) -\Sigma(\mathbf{q},\omega),
\end{equation}
where  $\Sigma(\mathbf{q},\omega)$ is the self-energy due to interactions. Its real part renormalizes $\hbar \omega(\mathbf{q}$), while its imaginary part is responsible for finite lifetime~\cite{Chernyshev2009,Zhitomirsky2013,Habel2024}.  The self-energy can be computed {\em perturbatively}, by evaluating selected Feynman diagrams containing loops~\cite{Chernyshev2009,Zhitomirsky2013,Habel2024}. Alternatively, one can compute the exact GF in Eq.~\eqref{eq:gf} for spin Hamiltonians defined on low-dimensional lattices {\em nonperturbatively} via numerically exact diagonalization~\cite{Bajpai2021} or (quasi)exact algorithms based on time-dependent matrix product states (TDMPS)~\cite{Paeckel2019}. These include~\footnote[2]{Note that comparison of  performance of tDMRG vs. TDVP methods for different quantum spin Hamiltonians can be found in Ref.~\cite{Chanda2020}} original time-dependent density matrix renormalization group (tDMRG)~\cite{White2004,Daley2004,Schmitteckert2004,Feiguin2011} and time-dependent variational principle (TDVP)~\cite{Haegeman2016,Chanda2020}. Instead of ``sharp bands'' of noninteracting magnons, one  then plots (as we do in Fig.~\ref{fig:fig1} obtained from TDMPS) either the spectral function~\cite{Chernyshev2009,Zhitomirsky2013,Bajpai2021} 
 \begin{equation}\label{eq:spectralf}
 A(\mathbf{q},\omega) = -\frac{1}{\pi} \mathrm{Im}\, G(\mathbf{q},\omega),
 \end{equation}
or the dynamical spin structure factor (DSSF), $S(\mathbf{q},\omega)$. The latter is directly measurable in neutron scattering experiments~\cite{Bayrakci2013}. The quantities $A(\mathbf{q},\omega)$ and $S(\mathbf{q},\omega)$ can also be related to each other~\cite{Mourigal2013}.

In particular, the very recent analyses~\cite{Gohlke2023} have revealed that LSWT producing ``sharp bands'' of noninteracting magnons can also introduce spurious symmetries, thereby leading to magnon 
spectrum that can be far away from the correct one obtained from quantum many-body calculations via tDMRG~\cite{Gohlke2023}. One reason for this discrepancy 
is generated when the Heisenberg effective spin Hamiltonian contains further than nearest-neighbor (NN) exchange interactions, and they are not equal to each other. Another one is when AFs contain very different exchange interactions~\cite{Gohlke2023} within two sublattices. This 
is precisely the situation (top row of panels of Fig.~\ref{fig:fig1}) encountered in RuO$_2$ as an early prototypical~\footnote[3]{Note that recent experimental~\cite{Hiraishi2024} and theoretical~\cite{Smolyanyuk2024} scrutiny finds RuO$_2$ 
to be actually nonmagnetic in bulk form, but it remains altermagnetic metal in few atomic layer form~\cite{Jeong2024}. Thus, these developments do not affect our study focused on  just a single layer in Fig.~\ref{fig:fig0}, as well as on general procedure for handling magnon-magnon interactions which can be applied to analyze thereby induced modification of chiral magnon bands of other experimentally confirmed~\cite{Liu2024} altermagnetic materials.} example of altermagnetic material~\cite{Smejkal2022,Smejkal2022a}. These newly established class of magnets were originally considered~\cite{Hayami2019,Yuan2020,Smejkal2020,Mazin2021} as  just another type of collinear AFs. They have attracted considerable attention due to their unusual fundamental properties and potential applications in spintronics. Akin to  ferromagnets (FMs), they exhibit time-reversal symmetry breaking and  transport properties like the 
the anomalous Hall effect, tunneling magnetoresistance and magneto-optics~\cite{Smejkal2022,Smejkal2022a}. This is due to AF ordering-induced  momentum-dependent spin splitting of electronic bands, analogous to materials with SO-split electronic bands but without requiring noncentrosymmetric crystals and high atomic number elements underlying materials with large SO coupling~\cite{Yuan2020}. The ``sharp bands'' of noninteracting magnons in  RuO$_2$ have already been computed~\cite{Smejkal2023}, exhibiting chirality akin to FMs but with linear energy-momentum dispersion akin to AFs. The Landau damping~\cite{Bonetti2022,Buczek2009,TancogneDejean2020} of magnons in RuO$_2$ was predicted in Ref.~\cite{Smejkal2023} to be suppressed by the spin-split electronic structure of altermagnets, but subsequent studies~\cite{Costa2024} find it to be finite and highly anisotropic with ability to complete  suppress  magnon propagation along selected spatial
directions.  On the other hand, possible damping due to magnon-magnon interactions remains unexplored. 

In this Letter, we compute magnon spectrum of an altermagnetic spin Hamiltonian~\cite{Smejkal2023} of RuO$_2$  via three different methods. For reference, we first obtain chiral ``sharp bands''  of noninteracting magnons on 4-leg cylinder in Fig.~\ref{fig:fig0} using LSWT~\cite{Chudnovsky2006,Gohlke2023}. Our LSWT also reproduces chiral bands of RuO$_2$ on three-dimensional (3D) lattice from Ref.~\cite{Smejkal2023}, see Supplemental Material (SM)~\footnote[1]{Supplemental Material at \href{https://wiki.physics.udel.edu/qttg/Publications}{https://wiki.physics.udel.edu/qttg/Publications}, including  Refs.~\cite{Berlijn2017, Sugiura2012, Sugiura2013, Gao2024, Yan2011, Press92, dosSantos2018}, provides: additional figure showing how our LSWT reproduces exactly  the ``sharp bands'' from Ref.~\cite{Smejkal2023} for $\mathrm{RuO}_2$ on its 3D lattice;  characterization of the GS of altermagnetic 4-leg cylinder in Fig.~\ref{fig:fig0}; further details of numerical procedures for obtaining spectral function via TDMPS; quantification of spontaneous magnon decay rates; and the role of magnetic anisotropy}. For the same  quasi-one-dimensional 4-leg cylinder, we then compute magnon spectrum encoded by $A(q_x,\omega)$ [Eq.~\eqref{eq:spectralf}] as obtained via nonpertrubative quantum many-body calculations based on TDVP-implemented algorithm~\cite{White2004} for magnonic retarded GF [Eq.~\eqref{eq:dmrggf}]. Finally, we extract magnon spectrum from classical atomistic spin dynamics (ASD)~\cite{Evans2014, Skubic2008, Etz2015} by calculating~\cite{Etz2015, Skubic2008} the DSSF [Eq.~\eqref{eq:dsf}] from stochastic Landau-Lifshitz-Gilbert (LLG) equation-based simulations of the same 4-leg cylinder but hosting classical spin vector on each site. The ASD simulations can take into account both magnon-magnon interactions~\cite{Zheng2023a} (as the LLG equation is nonlinear) and finite temperature (via noise terms in the LLG equation~\cite{Evans2014}, but only at the {\em classical} level. Our principal results for quantum-mechanically treated noninteracting [red and orange lines in Fig.~\ref{fig:fig1}(a),(e)] and interacting [blue traces in Fig.~\ref{fig:fig1}(a),(e)] magnons are 
shown in Fig.~\ref{fig:fig1}. In the same Figure, the classical treatment of interacting altermagnetic magnons via ASD is shown in Fig.~\ref{fig:fig1}(d),(h). Prior to discussing these results, we introduce useful concepts and notation.

\textit{Models and Methods.}---We employ a Heisenberg Hamiltonian~\cite{Szilva2023, Smejkal2023}
\begin{equation}\label{eq:hamiltonian}
\hat{H} = J \sum_{\langle i j \rangle} \hat{\mathbf{S}}_i \cdot \hat{\mathbf{S}}_j+\sum_{\langle \langle ij \rangle \rangle} J_{ij} \hat{\mathbf{S}}_i\cdot\hat{\mathbf{S}}_j,
\end{equation}
on a 4-leg cylinder with periodic boundary conditions in the transverse direction [Fig.~\ref{fig:fig0}]. Here  $\hat{\mathbf{S}}_i$ is the operator of $S=1/2$ spin localized at a site $i$ of the lattice; $\langle ij \rangle$  denotes the summation over NN sites; and  $\langle \langle ij \rangle \rangle$ denotes the summation over next-NN sites. Note that the exchange coupling $J_{ij}$ is direction-dependent and only takes two possible values $J_{A}$ and $J_B$, thereby, taking into account altermagnetic electronic spectrum effects on the effective spin Hamiltonian. The specific values of exchange interactions, \mbox{$J_A=1.25$ meV} and \mbox{$J_B=-0.6$ meV}, we use were extracted from DFT calculations of electronic spectrum of RuO$_2$ in Ref.~\cite{Smejkal2023}. Note that $J_A \neq J_B$ is direct consequence of spin-splitting of electronic spectrum of altermagnets (for more details, see the SM~\footnotemark[1]).

The LSWT spectrum of noninteracting magnon in AFs is   derived standardly~\cite{Chudnovsky2006, Gohlke2023}
from HP-transformed~\cite{Primakoff1940} effective spin Hamiltonian [Eq.~\eqref{eq:hamiltonian}] in terms of bosonic operators
\begin{subequations}
\begin{align}
    \hat{S}^z_{i,A} = \frac{1}{2}-\hat{a}^\dagger_i \hat{a}_i&, \label{eq:deviation} \: \hat{S}^z_{i,B} = \hat{b}^\dagger_i \hat{b}_i -\frac{1}{2},\\
    \hat{S}^-_{i,A} = \sqrt{1-\hat{a}^\dagger_i \hat{a}_i} \hat{a}_i&,  \: \hat{S}^-_{i,B}=\hat{b}_i^\dagger \sqrt{1-\hat{b}_i^\dagger \hat{b}_i},\\
    \hat{S}^+_{i,A} = \hat{a}^\dagger_i \sqrt{1-\hat{a}^\dagger_i \hat{a}_i} &, \:\hat{S}^+_{i,B}=\sqrt{1-\hat{b}_i^\dagger \hat{b}_i}\hat{b}_i,
\end{align}
\end{subequations}
after such transformations are {\em linearized}---$\hat{S}^z_{i,A(B)} =\frac{1}{2}-\hat{a}^\dagger_i \hat{a}_i (-\frac{1}{2}+\hat{b}^\dagger_i \hat{b}_i)$, $\hat{S}^-_{i,A(B)}=\hat{a}_i(\hat{b}_i^\dagger)$, and $\hat{S}^+_{i,A(B)}=\hat{a}_i^\dagger (\hat{b}_i)$---and the resulting HP Hamiltonian  truncated~\cite{Bajpai2021, Zheng2023a} to retain only quadratic terms in bosonic operators. In other words, one performs expansion~\cite{Chernyshev2009, Zhitomirsky2013} in $1/S$, starting with 
the order $1/S^2$, and retains only terms up to the order $(1/S)^{-1} = S$. Here, $\hat{a}_i(\hat{a}_i^\dagger)$ and  $\hat{b}_i(\hat{b}_i^\dagger)$  destroy (create) a boson at site $i$ within sublattice $A$ or $B$, respectively. Equation~\eqref{eq:deviation} clarifies that number of such bosonic excitations at a site $i$, as described by operators $\hat{a}_i^\dagger\hat{a}_i$ and $\hat{b}_i^\dagger\hat{b}_i$, captures the deviation of magnetic quantum number from its maximum value $S=1/2$. The same strategy leading to quadratic, and, therefore, exactly diagonalizable Hamiltonian can be applied to altermagnets. For this purpose and akin to the case of AFs, one applies the textbook~\cite{Chudnovsky2006} Bogoliubov-Valatin transformation of  $\hat{a}_i (\hat{a}_i^\dagger)$ and $\hat{b}_i (\hat{b}_i^\dagger)$ into $\hat{\alpha}_i (\hat{\alpha}_i^\dagger)$ and $\hat{\beta}_i (\hat{\beta}_i^\dagger)$. From thus produced Hamiltonian (displayed explicitly in the SM~\footnotemark[1]), that is exactly diagonalized by Bogoliubov-Valatin transformation, we read of the energy-momentum dispersion of its two  {\em non-degenerate} magnon bands
\begin{equation}\label{eq:HP}
    \hbar\omega (\mathbf{q}) = \pm \frac{\hbar }{2} [\omega_A(\mathbf{q})-\omega_B(\mathbf{q})] +\frac{\hbar}{2}\sqrt{\omega_{AB}^2(\mathbf{q})-4\gamma(\mathbf{q})}.
\end{equation}
Here $\gamma(\mathbf{q})$ is the usual function~\cite{Chudnovsky2006} encoding the crystalline lattice properties and its spatial dimensionality. The ``sharp bands'' of Eq.~\eqref{eq:HP} are plotted in red and orange lines in  Fig.~\ref{fig:fig1}(a),(e) for the case of 4-leg cylinder lattice in Fig.~\ref{fig:fig0}  whose $\gamma(q_x, q_y)=J(\cos q_xa +\cos q_ya)$. We also use  \mbox{$\omega_{AB}(q_x)= \omega_A(q_x) + \omega_B(q_x)$}; $\omega_{A}(q_x, q_y) = J-J_A/2-J_B/2+[J_A\cos(q_x+q_y)a+J_B\cos(q_x-q_y)a]/2$; $\omega_{B}(q_x, q_y) = J-J_A/2-J3_B/2+[J_B\cos(q_x+q_y)a+J_A\cos(q_x-q_y)a]/2$; and $a$ is the lattice constant. 

All possible magnon-magnon interactions are described by an infinite series~\cite{Bajpai2021,Zheng2023a} (or finite, when such infinite series is properly resumed~\cite{Bajpai2021,Vogl2020,Koenig2021}) of higher order terms on the top of the lowest-order quadratic HP Hamiltonian. By using TDMPS, all such terms are automatically taken into account. We perform TDMPS calculations via particular TDVP algorithm~\cite{Paeckel2019,Haegeman2016,Mingru2020} implemented in {\tt ITensor} package~\cite{Fishman2022} on a 4-leg cylinder with four sites in the transverse direction and $N=80$ sites in the longitudinal direction. We follow the same scheme of Ref.~\cite{White2004} to obtain the numerically exact retarded GF
\begin{equation}\label{eq:dmrggf}
    G(x,t) = -i\hbar\langle \Psi_0 | \hat{S}_x^z(t) \hat{S}_c^z(0)|\Psi_0\rangle,
\end{equation}
and from it the spectral function via Eq.~\eqref{eq:spectralf}. Note that demonstration of how to obtain these two quantities was originally  demonstrated via tDMRG algorithm applied to quantum spin chains~\cite{White2004}.  The ground state (GS), $|\Psi_0\rangle$, is found using up to 30 sweeps and a maximum bond dimension $\chi = 200$, which guarantees a maximum error of $\mathcal{O}(10^{-10})$. Then a wavepacket is created by applying $\hat{S}_c^z|\Psi_0\rangle$~\cite{White2004}, where $c=40$ is the lattice site in the middle of the ladder.  The magnon spectral function in Eq.~\eqref{eq:spectralf} is then obtained as the Fourier transform of  Eq.~\eqref{eq:dmrggf} to frequency and momentum, where a Gaussian window function for the Fourier transform in time is typically employed~\cite{White2004}. 


Classical ASD simulations were performed on the same 4-leg cylinder [Fig.~\ref{fig:fig0}], but using classical version of Hamiltonian in Eq.~\eqref{eq:hamiltonian} where vectors of fixed length $\mathbf{S}_i(t)$ replace spin operators $\hat{\mathbf{S}}_i$ and are evolved by the stochastic  (to include temperature~\cite{Evans2014}) LLG equation~\cite{Evans2014, Skubic2008}. This makes it possible to compute the DSSF~\cite{Etz2015} 
\begin{equation}\label{eq:dsf}
\mathcal{S}^{\alpha \alpha} (\mathbf{q},\omega)=\frac{1}{\sqrt{2\pi}N} \sum_{\mathbf{r}_i,\mathbf{r}_j}e^{i\mathbf{q} \cdot (\mathbf{r}_i-\mathbf{r}_j)}\int\limits_{-\infty}^{\infty}dt\, e^{i\omega t}C^{\alpha \alpha}(\mathbf{r}_i-\mathbf{r}_j,t).
\end{equation}
Here $\alpha=x,y,z$; N is the number of atoms per cell; and $C^{\alpha \alpha}(\mathbf{r}_i-\mathbf{r}_j,t)=\overline{S^\alpha_i(t) S^\alpha_{j}(0)} - \overline{S^\alpha_i (t)} \ \overline{S^\alpha_{j}(0)}$ measures correlations between spins with $\overline{\cdots}$ denoting ensemble average. The ensemble averaging is performed over different time segments during which a system of coupled LLG equations is solved to obtain trajectories $\mathbf{S}_i(t)$. The DSSF is directly measurable in neutron scattering experiments on bulk materials, or it can be obtained from spin-polarized high resolution electron energy loss spectroscopy measurements on thin films~\cite{Etz2015}. By plotting $\mathcal{S}^{\alpha\alpha}(\mathbf{q},\omega)$ in reciprocal space and by identifying the energies where its peaks appear~\cite{Etz2015}, we extract the energy-momentum dispersion of magnons [Fig.~\ref{fig:fig1}(d),(h)], in complete analogy to how dispersion is extracted from experimental neutron scattering data.

\textit{Results and discussion.}---The ``sharp bands'' of noninteracting magnons within LSWT track broadened and shifted interacting magnon bands \textit{only} at the Brillouin zone (BZ) center or edges in Fig.~\ref{fig:fig1}(a), while deviating in other parts of BZ where broadening due to magnon-magnon interactions concurrently increases. Comparing Figs.~\ref{fig:fig1}(a) and ~\ref{fig:fig1}(e), where the latter employs identical values of two exchange interaction constants $J_A=J_B$ between NN spins within sublattices $A$ and $B$ of altermagnet, clarifies how such discrepancy arises from $J_A \neq J_B$. This finding adds yet another case into the atlas of effective spin Hamiltonians~\cite{Chernyshev2009,Zhitomirsky2013, Mourigal2013,Gohlke2023,Habel2024,Winter2017,Smit2020, Sourounis2024} that can cause failure of LSWT. A deeper insight into the microscopic origin of the discrepancy can be obtained by analyzing higher order $n>2$ terms $\hat{H}_\mathrm{HP}^{(n)}$ in the Taylor series expansion~\cite{Bajpai2021, Chernyshev2009, Zheng2023a} of HP-bosonized effective spin Hamiltonian [Eq.~\eqref{eq:hamiltonian}],  $\hat{H}_\mathrm{HP} = \hat{H}_\mathrm{HP}^{(2)} + \hat{H}_\mathrm{HP}^{(4)} + \mathcal{O}(1/S)$. Here, subscript $n$ denotes presence of $n$th power in bosonic operators. Note that all orders $\hat{H}_\mathrm{HP}^{(n)}$ are captured by numerically (quasi)exact TDMPS calculations, but terms $n>4$ are progressively smaller, so we focus on $n= 4$ term. Since $\hat{H}_\mathrm{HP}^{(3)}$, as the first correction to  $\hat{H}_\mathrm{HP}^{(2)}$ of LSWT, is {\em zero} (the same applies to all other terms of odd $n$ in collinear AFs~\cite{Chernyshev2009}), one would na\"ively expect~\cite{Chernyshev2009} that magnon-magnon interactions encoded by $\hat{H}_\mathrm{HP}^{(4)}$ provide only a small correction. However,  examples of frustrated AFs exist~\cite{Winter2017} where  $\hat{H}_\mathrm{HP}^{(4)}$ can open a channel for large decay rate of magnons. Since $J_A \neq J_B$ is also a source of frustration, we analyze in Fig.~\ref{fig:fig1}(b),(f) if one-magnon spectrum of LSWT is energetically degenerate with three-magnon continuum in which case $\hat{H}_\mathrm{HP}^{(4)} = \sum_{q_1,q_2,q_3}V_{q_1,q_2,q_3} \hat{\alpha}_{q_1}^\dagger \hat{\alpha}_{q_2}^\dagger \hat{\alpha}_{q_3}^\dagger \hat{\beta}_{q_1+q_2+q_3}+ (\hat{\alpha}\leftrightarrow \hat{\beta})+\mathrm{H.c.}$, for the case of altermagnet in the focus of our study, can become effective. Figure~\ref{fig:fig1}(b) shows that the upper LSWT band overlaps with the three-magnon continuum, meaning that $\hat{H}_\mathrm{HP}^{(4)}$ will open a channel for decay of LSWT magnons into three-magnon states. The broadening of lower bands in Figs.~\ref{fig:fig1}(a) and \ref{fig:fig1}(e), despite {\em lack of degeneracy} between it and the three-magnon continuum [Fig.~\ref{fig:fig1}(b)], is actually an artifact of the window function employed when Fourier transforming Eq.~\eqref{eq:dmrggf} into Eq.~\eqref{eq:spectralf}. On the other hand, broadening of upper bands in Figs.~\ref{fig:fig1}(a) and \ref{fig:fig1}(e) is a genuine consequence of magnon-magnon interactions encoded by $H_{\mathrm{HP}}^{(4)}$. While this term also shifts upper band away from LSWT upper band in Fig.~\ref{fig:fig1}(a), upper band in Fig.~\ref{fig:fig1}(e) is only broadened without being renormalized. Note also that the upper band in Fig.~\ref{fig:fig1}(a) being shifted downward is the consequence of stronger hybridization on low-dimensional lattices~\cite{Verresen2019}, while on the 3D lattice of RuO$_2$ we expect band renormalization effects to be smaller.

The success of LLG equation in numerical modeling, via classical micromagnetics~\cite{Berkov2008} or ASD~\cite{Evans2014, Skubic2008, Etz2015, uppasd}, of local magnetization within FMs has also motivated its application to AFs~\cite{Cheng2014a,Li2020b,Moreels2024} and very recently to altermagnets~\cite{Weissenhofer2024a}. In the case of magnon spectrum calculation, ASD provides a computationally inexpensive (when compared to TDMPS) route that can easily include finite temperature effects via stochastic terms in the LLG equation~\cite{Skubic2008,Etz2015}. Due to nonlinearity of LLG equation, magnon-magnon interactions are intrinsically taken  into account. The ASD-computed magnon spectrum in Fig.~\ref{fig:fig1}(d),(h) exhibits bands that are similar to LSWT bands, but broadened even at \mbox{$T=1$ K} and using very small Gilbert damping $\alpha=10^{-4}$ (larger damping would trivially increase broadening). Thus observed broadening independently confirms the importance of magnon-magnon interactions for the effective spin Hamiltonian of RuO$_2$ in Eq.~\eqref{eq:hamiltonian}, which is not simply an artifact of low-dimensional lattice required for TDMPS calculations. However, ASD broadened bands do not trace TDMPS calculations [Fig.~\ref{fig:fig1}(a) vs. ~\ref{fig:fig1}(d)] as  magnon-magnon interactions captured quantum-mechanically  are shifting TDMPS bands away from either LSWT or ASD bands, as well as broadening them more than in ASD.  Such failure of ASD to capture TDMPS-computed magnon spectrum is not surprising, as it is well-known that AFs harbor entanglement~\cite{Song2011,Kamra2020} in their GS (leading to $\langle \Psi_0|\hat{\mathbf{S}}_i |\Psi_0\rangle=0$ in both AFs and altermagnets) and excited states.  Any residual  entanglement in quantum states of magnons will make  
their classical description via the LLG equation {\em inapplicable}~\cite{GarciaGaitan2024a}. While it is na\"ively expected that finite temperature and large number of spins will diminish entanglement of macroscopically large number of spins, it is precisely that such entanglement has been detected very recently in the GS of AFs via neutron scattering experiments~\cite{Scheie2021} up to surprisingly high temperature $T \lesssim 200$ K. Furthermore, when AFs are pushed out of equilibrium, even stronger entanglement is produced~\cite{GarciaGaitan2024a} that can remain nonzero~\cite{GarciaGaitan2024a}  (especially for spin $S=1/2$ and 
$S=1$) despite interactions with dissipative environment. 

{\em A litmus test for the importance of magnon-magnon interactions via Raman scattering experiments}.---Raman scattering (RS), based on inelastic scattering of light~\cite{Devereaux2007,Lemmens2003,Fleury1968,Cottam1972a,Hutchings1970} where the kinetic energy of an incident photon is increased (Stokes RS) or reduced (anti-Stokes RS), offers a unique probe of collective excitations in solids. For example, RS has been applied both to magnons or phonons, where one also has to differentiate~\cite{Hutchings1970,Fleury1968} their respective signals. Although RS probing of magnons lacks momentum dependence of neutron scattering~\cite{Bayrakci2013, Scheie2021}, it provides {\em high energy} resolution. In particular, two-magnon RS spectra reflect~\cite{Fleury1968,Davies1971} one-magnon DOS, thereby offering the following   experimental RS-based litmus test for the importance of magnon-magnon interactions: use one-magnon dispersion of LSWT of Ref.~\cite{Smejkal2023} to obtain DOS with rescaled (by a factor of two) energy scale $\rightarrow$ measure two-magnon RS spectra experimentally and plot together with DOS obtained in previous step $\rightarrow$ any discrepancy between these two curves signal the importance of magnon-magnon interactions. Figure~5 in Ref.~\cite{Prosnikov2021} givens an example of implementation of such test, and our Fig.~\ref{fig:fig1}(c) shows clear distinction between LSWT- and TDMPS-computed magnon DOS. 

{\em Conclusions.}---Comparing {\em three distinct} calculations of magnon spectrum of  RuO$_2$ altermagnet---quantum one-body via LSWT; quantum many-body via TDMPS; and classical many-body via ASD---divulges the following: ({\em i}) TDMPS bands of {\em interacting} magnons are shifted and broadened with respect to standard LSWT ``sharp bands''~\cite{Smejkal2023} of noninteracting magnons, as the signature of the relevance of magnon-magnon interactions where one magnon states can decay into three-magnon continuum; ({\em ii}) the distinction between DOS corresponding to these two spectra can be used to confirm the importance of magnon-magnon interactions, e.g., recently obtained~\cite{Smejkal2023}  magnon spectrum of RuO$_2$ (on its 3D lattice) and the corresponding DOS will not be able to explain two-magnon Raman scattering spectra if such interactions are indeed important for RuO$_2$; ({\em iii})  
classical ASD calculations of magnon spectra~\cite{Etz2015}, which are much less computationally expensive and highly popular~\cite{Cheng2014a, Li2020b} even for AFs, cannot trace quantum many-body calculations due to entanglement as highly nonclassical effect that AFs and altermagnets often  harbor~\cite{Song2011,Kamra2020,Scheie2021,GarciaGaitan2024a}  (in contrast, the success of LLG equation-based simulations of magnon spectra of FMs~\cite{Etz2015,Skubic2008} can be rigorously justified~\cite{GarciaGaitan2024a} by dissipative environment being able to completely wipe out their entanglement).

	\begin{acknowledgments}
		This research was supported by the US National Science Foundation (NSF) through the  University of Delaware Materials Research Science and Engineering Center, DMR-2011824. The supercomputing time was provided by DARWIN (Delaware Advanced Research Workforce and Innovation Network), which is supported by NSF Grant No. MRI-1919839.
        The paper originated from Research Projects Based Learning implemented within a graduate course {\em PHYS800: Advanced Graduate Seminar in Quantum Physics}~\cite{phys800} at the University of Delaware.
	\end{acknowledgments}

\bibliography{references}

\end{document}